\documentclass[showpacs,twocolumn,prb]{revtex4}

\usepackage{graphicx}
\usepackage{latexsym}
\usepackage{times}
\usepackage{psfrag,amsfonts}

\newcommand{\ket}[1]{| #1 \rangle}

\newcommand{\average}[1]{ \left\langle #1  \right\rangle}

\newcommand{\be}{\begin{equation}}
\newcommand{\ee}{\end{equation}}
\newcommand{\beq}{\begin{equation}}
\newcommand{\eeq}{\end{equation}}
\newcommand{\bae}{\begin{eqnarray}}
\newcommand{\eae}{\end{eqnarray}}

\def\CC{{\rm\kern.24em \vrule width.04em height1.46ex depth-.07ex
    \kern-.30em C}}
\def\P{{\rm I\kern-.25em P}}

\def\bbbc{{\mathchoice {\setbox0=\hbox{$\displaystyle\rm C$}\hbox{\hbox
to0pt{\kern0.4\wd0\vrule height0.9\ht0\hss}\box0}}
{\setbox0=\hbox{$\textstyle\rm C$}\hbox{\hbox
to0pt{\kern0.4\wd0\vrule height0.9\ht0\hss}\box0}}
{\setbox0=\hbox{$\scriptstyle\rm C$}\hbox{\hbox
to0pt{\kern0.4\wd0\vrule height0.9\ht0\hss}\box0}}
{\setbox0=\hbox{$\scriptscriptstyle\rm C$}\hbox{\hbox
to0pt{\kern0.4\wd0\vrule height0.9\ht0\hss}\box0}}}}

\def\bbbz{{\mathchoice {\hbox{$\sf\textstyle Z\kern-0.4em Z$}}
{\hbox{$\sf\textstyle Z\kern-0.4em Z$}} {\hbox{$\sf\scriptstyle
Z\kern-0.3em Z$}} {\hbox{$\sf\scriptscriptstyle Z\kern-0.2em Z$}}}}

\begin{document}

\title{Momentum-space analysis of multipartite entanglement at quantum phase transitions}

\author{Alberto Anfossi$^{1}$, Paolo Giorda$^{2}$ and Arianna Montorsi$^3$}

\address{$^1$Dipartimento di Fisica, Universit\`{a} di Bologna, viale C. Berti-Pichat 6/2, I-40127 Bologna, Italy}

\address{$^2$Institute for Scientific Interchange (ISI), Villa Gualino, Viale Settimio Severo 65, I-10133 Torino, Italy}

\address{$^3$Dipartimento di Fisica del Politecnico and CNISM, Corso Duca degli Abruzzi 24, I-10129 Torino, Italy}

\begin{abstract}

We investigate entanglement properties at quantum phase transitions of an integrable extended Hubbard model in the momentum space representation. Two elementary subsystems are recognized: the single mode of an electron, and the pair of modes (electrons coupled through the $\eta$-pairing mechanism). We first detect the two/multi-partite nature of each quantum phase transition by a comparative study of the singularities of Von Neumann entropy and quantum mutual information. We establish the existing relations between the correlations in the momentum representation and those exhibited in the complementary picture: the direct lattice representation. The presence of multipartite entanglement is then investigated in detail through the Q-measure, namely a generalization of the Meyer-Wallach measure of entanglement. Such a measure becomes increasingly sensitive to correlations of a multipartite nature increasing the size of the reduced density matrix. In momentum space, we succeed in obtaining the latter for our system at arbitrary size and we relate its behaviour to the nature of the various QPTs.

\end{abstract}


\maketitle

\section{Introduction}

The idea of studying and characterizing Quantum Phase Transitions (QPTs) by means of tools and concepts borrowed from the quantum information theory has provided in the past years a significant amount of interesting results \cite{amicoRMP}. In particular, the notion of entanglement and its measures is of central importance within this framework. Indeed, for several and different models in condensed matter physics it has been shown that partial derivatives of appropriate measures of entanglement with respect to the field driving the QPT keep track of the underlying critical phenomenon by developing a singularity \cite{fazio,OSNI,LatVidKor,WSL,VerER,AmVerETr,GDLL,AGMT,ADMO,LAJO,CGZ,AGM}. Moreover, from an entanglement study of QPTs one can gain further information such as the order or the scaling exponents as well as a deeper insight on the structure of the correlations involved. For instance in \cite{AGMT} it has been argued how, for an extended Hubbard model, a comparison of the Von Neumann entropy --a measure of quantum correlation-- of a single site with the quantum mutual information, measuring the total correlations between two sites, allows one to infer which kind of correlation (two-point or multipartite) plays the prominent role at a given quantum transition. Such a scheme has been  tested in \cite{AGM,Tribedi-Bose}. In particular,  in  \cite{AGM}  the two-point QPTs have been investigated by means of punctual measures of two-point quantum correlations.
As for the multipartite ones, the analysis becomes harder. In fact, the evaluation of the known direct measures of multipartite correlations requires either some optimization process that turns out to be in practice unfeasible, or the knowledge of the reduced density matrix relative to arbitrarily large subsystems.

Entanglement is in general a notion that depends on the choice of the subsystems whose correlations one is interested in analyzing. In particular, the same Hilbert space ${\cal{H}}$ can support different tensor product structures  and thus different choices of subsystems are possible \cite{zan virtualsub}. In describing many-body condensed matter systems one is naturally led to consider at least two different structures: the one induced by the real-space topology of the system under consideration, i.e., the {\em direct lattice picture}, and the one corresponding to the momentum space, i.e., the {\em reciprocal lattice picture}. Although the literature on entanglement in condensed matter has privileged the first description, one expects that the information obtained by the analysis of entanglement properties of elementary subsystems of the direct lattice (subsets of single sites of the ambient lattice) can be fruitfully integrated with that obtained by considering as elementary subsystems the momenta modes of the reciprocal lattice.

In this paper we provide the momentum space description of the ground state entanglement properties of the extended Hubbard model discussed in \cite{AAS,AGMT,AGM}. The study  in momentum space allows for a different and rich picture of the structure of correlations and of their relations with QPTs. Indeed, the relevance itself of multipartite (multi-mode) and two-point (two-mode) correlations at a given QPT can significantly vary when the above change of perspective is performed. Furthermore, the comparison of the (quantum) correlations in the momentum and direct lattice picture turns out to be a fruitful method of analysis. Indeed, this procedure allows one to enlighten the origin of the off-diagonal long-range order (ODLRO) displayed by the ground state for appropriate values of the parameters: this is directly related to the value of negativity between modes k and -k \cite{GA}.

Moreover, the possibility of determining the reduced density matrices of arbitrary subsystems composed by sets of momentum modes, allows one to overcome the above mentioned difficulties in measuring directly multipartite correlations. We exploit such   possibility to compute the generalization of the Meyer-Wallach measure of multipartite entanglement \cite{MeyWalQ,ScottQ}. We thus show that by increasing the subsystem size the partial derivative of $Q$ develops a singularity in correspondence with precisely those QPTs at which the dominant correlations were conjectured to be of multipartite nature.

The paper is organized as follows. In section \ref{Sec:model} we describe the phase diagram of the model under investigation and we introduce the momentum space description its ground state in the various phases. In section \ref{sec:results_general} we first discuss the general structure of the correlations in the ground state in the whole phase diagram; we related the picture emerging in the momentum space with the one previously obtained in the direct lattice picture; we discuss the role of the correlations at QPTs and we evaluate and discuss the Meyer-Wallach measure of multipartite entanglement. In section \ref{sec:conclusions} we summarize our conclusions.

\section{The bond-charge extended Hubbard model}
\label{Sec:model}

The model we investigate is the bond-charge extended Hubbard model for a particular value of the bond-charge interaction at which the model becomes integrable \cite{AAS}. In units of the nearest-neighbors hopping amplitude, its Hamiltonian reads:
\begin{equation}\label{ham_bc}
    H=- \sum_{<i,j>\sigma}[1 - (n_{i \bar{\sigma}}+n_{j \bar{\sigma}})] c_{i \sigma}^\dagger c_{j \sigma}+u \sum_i n_{i \uparrow}n_{i \downarrow}\, ,
\end{equation}
where $c_{{i} \sigma}^\dagger$ and $c_{{i} \sigma}^{} \,$ are fermionic creation and annihilation operators on a one-dimensional periodic chain of length $L$; $\sigma =\{ \uparrow, \downarrow \}$ is the spin label, $\bar{\sigma}$ denotes its opposite, ${n}^{}_{j \sigma} = c_{j \sigma}^\dagger c_{j \sigma}^{}$ is the spin-$\sigma$ electron charge, and $\langle {i} , \, {j} \rangle$ stands for neighboring sites on the chain; $u$ is dimensionless on-site Coulomb repulsion.

The two following feature make Hamiltonian (\ref{ham_bc}) integrable \cite{AAS}: $i)$ the number of doubly occupied sites is a conserved quantity; $ii)$ the role of spin orientation is irrelevant. For an open chain, any sequence of spins in the chain cannot be altered by $H_{BC}$, while for periodic boundary conditions, only the sequences of spins related by a cyclic permutation can be obtained. In particular, the ground state turns out to be degenerate with the fully polarized state.

\subsection{Eigenstates, eigenvalues and ground-state phase diagram} \label{Sec:phdiag}

The physics of the system described by $H$ is basically that of $N_s= \langle\hat{N_s}\rangle$ spinless fermions moving in a background of $L-N_s$ bosons, of which $N_d = \langle \hat{N}_d \rangle$ are pairs of particles and the remaining are empty sites. Both $N_s$ and $N_d$ are conserved quantities (and determine the total number of electrons $N=N_s + 2 N_d$) and the following ans\"{a}tz for the form of the  eigenstates can be formulated in momentum space:
\begin{equation}
    \ket{\psi (N_s,N_d)}= \mathcal{N} (\tilde{\eta}^\dagger)^{N_d}
      a^\dagger_{k_1\sigma_{k_1}}\dots a^\dagger_{k_{N_s}\sigma_{k_{N_s}}}\ket{\mbox{vac}} \; .\label{psi}
\end{equation}
Here $\mathcal{N}=\Bigl[(L-N_s-N_d)!/ (L-N_s)!N_d!\Bigr]^{1/2}$ is the normalization factor and $a^\dagger_{k_j\sigma}\doteq\frac{1}{\sqrt{L}}\sum_{l} e^{i\frac{2\pi}{L}lj}c_{l\sigma}^\dagger$.
The operator $\eta^\dagger =\sum_{j} a_{-k_j\uparrow}^\dagger a_{k_j\downarrow}^\dagger$ is the so-called eta operator, its action being to create a $0$-momentum couple of fermions delocalized over the reciprocal lattice.

\begin{figure}[!h]
        \includegraphics[height=10cm, width=10cm, keepaspectratio, viewport= 20 10 800 500,clip]{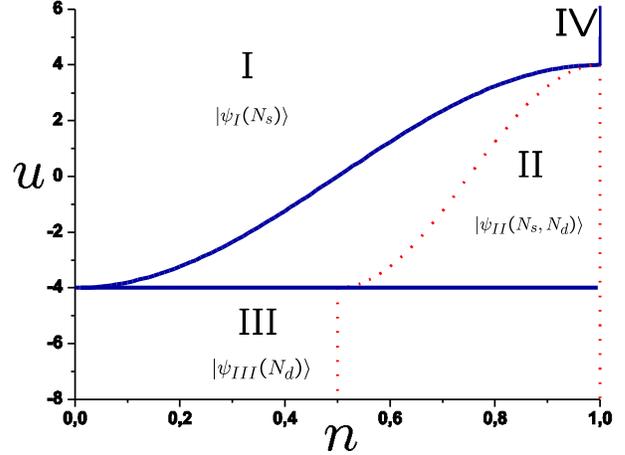}
        \caption{$k$-space representation of the ground-state phase diagram of $H$ in the $n$-$u$ plane. Dotted lines represent some cases of \emph{iso-correlation} curves, see sec. \ref{sec:results_entqpt}.}\label{fig:phdiag}
\end{figure}

The actual ground state $\ket{\psi_{GS}(N_s,N_d)}$ is chosen among the eigenstates (\ref{psi}) by requiring that $N_s$ and $N_d$ minimize the corresponding eigenvalue, which in the thermodynamic limit reads ($n_s=N_s/L$, $n_d=N_d/L$):
\begin{equation}
    e(n_s,n_d) = -\frac{2}{\pi}\sin{(\pi n_s)} + u n_d \,.
\end{equation}
The minimization equations can be solved, thus obtaining the ground-state phase diagram shown in Fig. \ref{fig:phdiag} in the $n$-$u$ plane (with $n=N/L$ average per-site occupation number).
Three distinct regimes are recognized, depending on $n_s$, $n_d$. For sufficiently high $u$, $n_d=0$ characterizes the conducting region $I$, which, for $n=1$ becomes insulating (phase IV). In this case  $n_s=n$. Keeping $n$ fixed and decreasing $u<u_c(n)=-4\cos(\pi n)$ we pass to region II, where the pairing mechanism becomes convenient and the energy minimum is realized for $n_d\neq0$. By further decreasing $u$ we enter in region III, where pairs are energetically favored with respect to single fermions; the energy minimum is thus obtained imposing the condition $n_s=0$ (i.e. $n_d=n/2$). Although both phases I and II are characterized by a conducting behavior, they are qualitatively different since only in phase II the occurrence of off-diagonal long-range order (ODLRO) -- which also survive in phase III -- shows up \cite{AGM}.\\

%\textbf{At half-filling and $u>4$ a further phase (IV) can be identified. In momentum space phase IV is not distinguished from phase I, both being characterized by a factorized ground-state. Nevertheless phase IV is qualitatively different form phase I since it is insulating.}

The standard approach to the study of (quantum) phase transitions consists in investigating the behavior of the (ground state) energy density at critical points. Within this framework, all QPTs here studied are second order ones: they are signaled by non analyticities of the second order partial derivatives of the ground-state density energy. When $u=\pm 4$ two transitions take place: II$\rightarrow$IV and II$\rightarrow$III. Here the significant derivative is the one with respect to the field $u$. Indeed, it is the second derivative that diverges at the critical points $u_c=\pm 4$:
\begin{equation}
\partial_u^2 E_{gs}(u,0)= \frac{-1}{2\pi\sqrt{16-u^2}}\;.
\label{Energy Divergency II->IV,II}
\end{equation}
At the transition II$\rightarrow$I the QPTs are again signaled by the non analyticities of the second order partial derivatives, but --at variance with the previous case-- these simply exhibit a discontinuity and do not diverge at the transition points $u_c = -4 \cos(\pi n)$:
  \bae
    \lim_{u\rightarrow u_c^+} \partial^2_u E_{gs}(u,n) &=& 0\neq
    \frac{-1}{8\pi\sqrt{1-\cos^2{(n\pi)}}}=\nonumber \\
    &=&\lim_{u\rightarrow u_c^-} \partial^2_u E_{gs}(u,n)\;, \nonumber \\
    \lim_{n\rightarrow n_c^+} \partial^2_n E_{gs}(u,n) &=& 0\neq2\pi\cos{(n\pi)}=\nonumber \\ &=&\lim_{u\rightarrow n_c^-} \partial^2_n E_{gs}(u,n)\;.\nonumber \\ 
    \label{Energy Divergency II->I}
  \eae

As already mentioned, the QPTs above described were studied in \cite{AGM} in terms of (quantum) correlations between different subsystems in the direct lattice picture. There the natural subsystems were the single/pairs of {\em sites} of the lattice. In this picture, the principal measures of correlations taken into account were: $i)$ the Von Neumann entropy of the generic single site $l$, $\mathcal{S}_{l}$, that measures its total quantum correlations with the rest of the chain; $ii)$ the mutual information between two sites $l,m$, $\mathcal{I}_{l,m}$, that measures the total (quantum and classical) correlations between the two sites; $iii)$ the negativity \cite{vidalvernerneg} (and concurrence \cite{Woo}) between the two sites $l,m$, $\mathcal{N}_{l,m}$, that measures only the quantum contribute to the correlations between the two sites.
Remarkably, the study of the derivatives of these measures of correlations allowed to distinguish the various QPTs in terms of relative relevance of the two-point (i.e., two-site) and multipartite correlations.
In particular, the divergence (\ref{Energy Divergency II->IV,II}) exhibited by the energy density is the same shown by the first derivative of both quantum mutual information and Von Neumann entropy at both transitions. This is to understand as a signature of the relevance of two-point entanglement at the transitions, relevance that was confirmed by the divergence of the \emph{entanglement ratio} that, indeed, originates from the spreading of the two-point quantum correlations along the whole chain.

In correspondence with transition II$\rightarrow$I --signaled by the discontinuities of the energy density (\ref{Energy Divergency II->I})-- while the partial derivative of the Von Neumann entropy still exhibits a logarithmic divergence, the partial derivative of the quantum mutual information is simply discontinuous. This is to understand as a manifestation of the relevance of multipartite entanglement at such transition.

\subsection{$k$-space representation}\label{sec:k_space}

The main aim of this paper is to extend the study of how the QPTs introduced in the previous sections can be described in terms of correlations in the $k$-space picture. The $k$-space representation of the eta-paring states with $N_d$ fixed and $N_s=0$ have been thoroughly analyzed in \cite{GA}. Here we first give a brief description of the main ingredients needed for the analysis; second we see how the picture can be extended to region II in which $N_s\neq0$, i.e., the ground state is filled at the same time by single fermions and pairs of particles.\\
As previously mentioned, each single eta operator in the the $k$-space representation can be written as:
\begin{equation}
 \eta^\dagger =\sum_{j}\hat{\eta}_{k_j}^\dagger  =\sum_{j} a_{-k_j\uparrow}^\dagger a_{k_j\downarrow}^\dagger\, .
\end{equation}
where $\hat{\eta}_{k_j}^\dagger = a_{-k_j\uparrow}^\dagger a_{k_j\downarrow}^\dagger $ and $a_{k_j\sigma}^\dagger,a_{k_j\sigma}$ are fermionic operators that create/annihilate a particle with given spin $\sigma_{k_j}$ in one of the $L$ momentum modes $k_j=2\pi j/L$ whose local basis is $\mathcal{B}_{k_j}=\{\ket{0}_{k_j}, \ket{\uparrow}_{k_j}, \ket{\downarrow}_{k_j}, \ket{\uparrow\downarrow}_{k_j}\}$.
First, let us notice that the different regions in Fig. \ref{fig:phdiag} correspond through (\ref{psi}) to different realizations of the ground states in $k$-space. In particular, in region $I$, at variance with the direct lattice picture, the ground state, being constituted by $N_s$ electron with definite momentum, is completely factorized:
\begin{equation}
    |\psi_{I}(N)\rangle=\otimes_{j=1}^N |\sigma_{k_j}\rangle\; . \label{psi1}
\end{equation}
In region II the ground state (\ref{psi}) is built by occupying with $N_s$ fermionic particles (with fixed spin $\sigma_{k_j}$) the modes of the reciprocal lattice
\bae
    k_j \in K_{k \leq k_{N_s}} &=& \{K_{\mbox{min}}=-\pi (N_s-1)/L,\dots, \nonumber\\
    && K_{\mbox{max}}=\pi (N_s-1)/L\}\;.
\eae
Since in region II part of the momentum space is occupied by $N_s$ single particles, the delocalization implemented by the eta operators can only take place  over a subset of momenta:
\begin{eqnarray*}
    K_{k>k_{N_s}}&=&\{\pi N_s/L, \dots,\pi \} \\
    &&\cup \{ -\pi+2\pi/L,\dots,-2\pi(N_s/2+1)/L \}.
\end{eqnarray*}
The number of distinct values of momenta available for the creation of pairs of particles, i.e., the order of $K_{k>k_{N_s}}$, is $L-N_s$. This fact can be represented by observing that the partition of set of possible $k$-modes in $K_{k \leq k_{N_s}}\cup K_{k>k_{N_s}}$ corresponds to a bipartition of the $k$-state space: ${\mathcal H}={\mathcal H}_{k \leq k_{N_s}} \otimes \mathcal{H}_{k>k_{N_s}}$. Considering the vacuum
\begin{equation}
\ket{\mbox{vac}}_{\mathcal H}=
\ket{\mbox{vac}}_{{\mathcal H}_{k \leq k_{N_s}}} \otimes \ket{\mbox{vac}}_{\mathcal{H}_{k>k_{N_s}}}
\end{equation}
the ground state in (\ref{psi}) can be written as
\begin{eqnarray*}
    \ket{\psi_{II}(N_s,N_d)} &\doteq& \mathcal{N}(\eta^\dagger)^{N_d}
    a^\dagger_{k_1\sigma}\dots a^\dagger_{k_{N_s}\sigma} \ket{\mbox{vac}}_{\mathcal H}\\
     &=& (\eta^\dagger )^{N_d} \ket{\mbox{vac}}_{\mathcal{H}_{k>k_{N_s}}} \ket{\psi_{I}(N_s)} \; .
\end{eqnarray*}

>From the point of view of the delocalization implemented by the eta-operators, region III represents the limiting case where the $N_d$ pairs can be created over the full momentum space:
\begin{equation}
    \ket{\psi_{III}(N_d)} \doteq \mathcal{N}(\eta^\dagger)^{N_d}
     \ket{\mbox{vac}}_{\mathcal H}
\end{equation}

The above discussion about region II implies that the only modes that are non-trivially correlated are those belonging to $\mathcal{H}_{k>k_{N_s}}$, while the modes in $\mathcal{H}_{k \leq k_{N_s}}$ are simply in a tensor product state where each mode is singly occupied by a spin-$\sigma$ particle. The characterization of region II and of the transitions II$\rightarrow$III,I can thus be done in terms of the correlations between the modes over which the pairs of particles created by the eta-operators are delocalized. The influence of the single particles in this picture is to reduce of the space available for the delocalization and this reduction has an important role in determining many properties of the state correlations.

\section{Results in momentum representation}\label{sec:results_general}

We now generalize the results obtained in \cite{GA} for the eta-pairing states with $N_s=0$ to the generic states (\ref{psi}) where the eta operators act on a reference state characterized by the presence of single particles with a given momentum. The extension allows one to investigate, in $k$-space, the behavior of the correlations at the QPTs exhibited by the model (\ref{ham_bc}).

\subsection{Local structure of correlations in k-space}\label{sec:results_correlations}

A first significant result in the analysis of the structure of correlations is the acknowledgement that both the generic single mode $k_j$ and the {\em paired modes} $(-k_j,k_j)$ are fundamental building blocks.  The basic objects that one has to evaluate in order to derive the appropriate measures of correlations are the reduced density matrices of the given subsystems: $i$) $\rho_{k_j}$ for the single mode; $ii)$  $\rho_{k_i,k_j}$ for two generic modes; $iii)$ $\rho_{(-k_i,k_i),(-k_j,k_j)}$ for two pairs of modes.\\
The general considerations that allow one to determine the structure of these density matrices are the following. Since the state $(\eta^\dagger)^{N_d}\ket{\mbox{vac}}_{\mathcal{H}_{k>k_{N_s}}}$  is built by the repeated application of operators $\eta^\dagger_{k_j}$, each time the $-k_j$ mode is occupied, the $k_j$ one must be occupied correspondingly. Therefore, for example, the only non-vanishing elements of $\rho_{k_j}$ are the diagonal ones. As far as $\rho_{k_i,k_j}$ ($\rho_{(-k_i,k_i),(-k_j,k_j)}$) is concerned, the only $d$-modes ($d=2$, $d=4$ respectively) projectors $P_{\nu_d}^d$ that have non-vanishing expectation values are those preserving the number $\nu_d=0,\dots,d$ of pairs of particles $(\sigma_{k_j},\bar{\sigma}_{-k_j})$. For a given $\nu_d$ the expectation values are all equal and in the TDL they read
\begin{equation}
    \langle P_\alpha^d\rangle=a^{\nu_d}(1-a)^{d-\nu_d},\; \mbox{ where  }\; a=\frac{n_d}{1-n_s}\;.\label{Projectors}
\end{equation}
It turns out that these calculations can be generalized to an arbitrary number of both single and paired modes, one can thus explicitly find the diagonal form of the reduced density matrix of any subsystem. In the TDL, \emph{the latter and consequently all measures of correlations depend exclusively on the parameter $a$}.\\
Some basic considerations are here in order. As previously observed, the presence of the $N_s$ single particles with fixed spin $\sigma_{k_j}$ ($k_j < k_{N_s}$) in region II has the effect of reducing the available modes over which the eta-operators can delocalize the pairs of particles they create.
This effect can be substantiated by considering the average number of particles with a given momentum $k_j$: $\average{n_{k_j}}=\average{n_{k_j\uparrow}+n_{k_j\downarrow}}$. The latter is simply constant and equal to one for all $k_j \in K_{k \leq k_{N_s}}$. For $K_{k>k_{N_s}}$ we have that $\average{n_{k_j}}=2\average{n_{k_j\sigma}} =2N_d/(L-N_s)=2 n_d/(1-n_s)=(n-n_s)/(1-n_s)=2a$. This means that, in general, the effect on correlations of the delocalization implemented by the eta-operators depends only on the density of particles, i.e., it is related to the filling $N=N_s+2 N_d$ and the number of available modes $|K_{k>k_{N_s}}|=L-N_s$.

We can now turn to the detailed description of the structure of correlations at a local level. In particular, the generic mode $k_j$ is quantum correlated with the rest of the system and, in the TDL, the Von Neumann entropy reads $\mathcal{S}_{k_j} = -2[a\log a+(1-a)\log{(1-a)}]$; furthermore $k_j$ has two-mode correlations only with the mode $-k_j$:
\bae
 \mathcal{I}_{k_i,k_j} &=& \left\{\begin{array}{ll}
  \mathcal{S}_{k_i}+2a(1-a) & k_i=-k_j\\
  0 & k_i\neq -k_j
 \end{array}\right.\label{MIkikj}\\
 \mathcal{N}_{k_i,k_j} &=& \left\{\begin{array}{ll}
  a(1-a)/3 & k_i=-k_j\\
  0 & k_i\neq -k_j
  \end{array}\right.
\eae
where $\mathcal{I}_{k_i,k_j}$ ($\mathcal{N}_{k_i,k_j}$) is the mutual information (negativity) between two generic modes $(k_i,k_j)$.

As for the paired modes $(-k_j,k_j)$, each of them exhibits quantum correlations only with the rest of the system as a whole, indeed, while their Von Neumann entropy reads
\begin{equation}
    \mathcal{S}_{k_i,k_j} = -2\left(a\log{a}+(1-a)\log{(1-a)}\right)+a(1-a)\;,\label{VNEkikj}
\end{equation}
the {\em two-pairs} correlations  $\mathcal{I}_{(-k_j,k_j),(-k_i,k_i)} = 2a(1-a)[2+ a(1-a)(3 \log_2{3} - 5)]$ are just of classical nature since
$\mathcal{N}_{(-k_j,k_j),(-k_i,k_i)} = 0$.

The above picture naturally suggests to investigate at QPTs the behavior of the correlations concerning the single mode and the paired modes.
As already anticipated the expression of $\average{n_{k_j}}$ is at the basis of the definition and the behaviour of (\ref{Projectors}-\ref{VNEkikj}) and in general of the correlations between any pair of subsystems (in the limit of large $L$). Therefore,
the derivative of the measures of correlations at the various phase transitions can be expressed in terms of the derivatives of $\average{n_{k_j}}$, and this turns out to have important consequences for the discrimination of the various QPTs.

The above description of the network of correlations in $k$-space can be fruitfully completed by analyzing the relations between such kind of correlations and some meaningful quantities that are defined in the direct lattice picture. In particular one can see how ODLRO, which is defined in the direct lattice, is in fact directly related to negativity between paired modes, generalizing the relation established in \cite{GA}. Indeed, in this model the ODLRO is defined as
\be
\lim_{|i-j|\rightarrow \infty} \average{\eta^\dagger_i \eta_j}= n_d(1-n_d)\propto \mathcal{I}_{i,j}^{III}\label{Eq: ODLRO Reg III}
\ee
where  $\eta_i^\dagger = c_{i,\uparrow}^\dagger c_{i,\downarrow}^\dagger$  is the operator that creates a pair of particles at the lattice site $i$; the the proportionality to the quantum mutual information between two sites $i,j$ evaluated in region III i.e., $\mathcal{I}_{i,j}^{III}$,  was found in \cite{AGM}. Here we find that
\bae
\lim_{|i-j|\rightarrow \infty} \average{\eta^\dagger_i \eta_j}&=& n_d(1-n_s-n_d)\propto \mathcal{N}_{k_i,-k_i}(1-n_s)^2.\nonumber\\
\label{Eq: ODLRO vs Neg}
\eae
which means that, even in region II i.e., where $n_s\neq0$, the negativity and the ODRLO are directly related and the latter relation reduces to the one found in \cite{GA} when evaluated in region III $(n_s=0)$.\\
The relation between quantities defined in the different pictures can be further extended by considering the full expression of $\mathcal{I}_{i,j}^{II}$ i.e., the quantum mutual information between two sites $i,j$ in region II. We recall that in \cite{AGM} we derived the following expression
\be
\mathcal{I}_{|i-j|}^{II}=\mathcal{I}_{|i-j|}^{I}+\mathcal{I}_{\infty}^{II},
\label{Eq: Mutual_II_I_infty}
\ee
where $\mathcal{I}_{|i-j|}^{I}$ is the mutual information between the site $i,j$ evaluated in region I ($n_d=0,\, n=n_s$) and $\mathcal{I}_{\infty}^{II}=\lim_{|i-j|\rightarrow \infty} \mathcal{I}_{|i-j|}^{II}$ is the mutual information between two infinitely distant sites in region II. This equation suggests that in region II one can distinguish the part of the direct lattice {\em two-sites} correlations i.e., $\mathcal{I}_{|i-j|}^{I}$ which are determined by the fraction of $n_s$ single particles which are not correlated in $k$-space,  by another contribution, $\mathcal{I}_{\infty}^{II}$, which takes into account, in direct lattice, the effect of the eta paring mechanism with which the rest of the fraction of particles $n_d$ are created . Here we can now fully understand this peculiar behaviour in terms of quantum correlations between paired modes. Indeed on one hand, by using the elements of the two-sites reduced density matrix which were analytically evaluated in \cite{AGMT,AGM}, one can show that the proportionality established in (\ref{Eq: ODLRO Reg III}) can be in some sense extended in region II: we have that in this region the ODLRO coincides with $\mathcal{I}_{\infty}^{II}$. On the other hand, thanks to (\ref{Eq: ODLRO vs Neg}), this quantity can be directly linked to the $\mathcal{N}_{k_i,-k_i}$ and this roots the origin of this part of the {\em two sites} correlations in the existing correlations between {\em paired momentum modes}. \\
It turns out that this is important relation between the (quantum) correlations pertaining the two complementary pictures, will also be relevant in the next section when we discuss the behaviour of the various correlation measures at the QPTs.

\subsection{Entanglement and quantum phase transitions} \label{sec:results_entqpt}

In previous papers \cite{AGMT,AGM} we showed how, in the direct-lattice picture, the transition II$\rightarrow$I  (see Fig. \ref{fig:phdiag}) can be distinguished from transitions  II$\rightarrow$III and and I$\rightarrow$IV on the basis of the different kind of correlations (two-site versus multipartite) among different subsystems. Such a distinction between these QPTs can be achieved in the $k$-space picture only in a different way. In fact, we first note that at variance with the direct-lattice picture, here the insulating state which represents the ground-state at half-filling for $u>4$ (insulating phase IV), as far as correlations are concerned, is not qualitatively distinct from the conducting state (\ref{psi1}) which occurs at arbitrary filling when $u\ge -4 \cos(\pi n)$. Indeed, in $k$-space the conducting state (\ref{psi1}) is fully factorized for any filling $n$, so that entanglement is identically zero at any $n$ and the transition I$\rightarrow$IV cannot be detected from entanglement analysis.
In the ambient lattice the factorization property holds at half-filling only.
\begin{figure}[!h]
\begin{centering}
\fbox{\includegraphics[height=7cm, width=6cm, keepaspectratio, viewport= 0 70 680
520, clip]{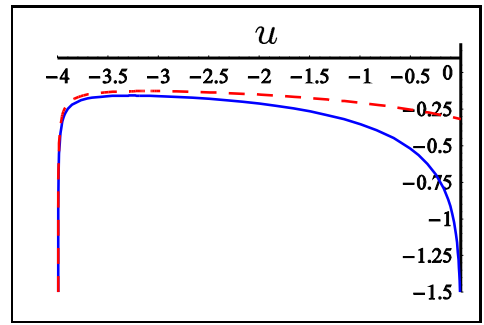}}
 \caption{(Color online) Plot of $\partial_u\mathcal{S}_{-k_j,k_j}$ (straight blue line) and $\partial_u\mathcal{I}_{(-k_i,k_i),(-k_j,k_j)}$ (dashed red line) in region II. }
 \label{meas-qpts}
 \end{centering}
\end{figure}

\begin{figure}[!h]
\begin{centering}
\fbox{\includegraphics[height=7cm, width=6cm,  keepaspectratio,viewport= 0 40 680
500, clip]{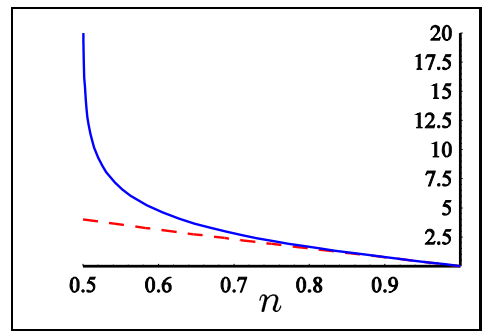}}
 \caption{(Color online) Plot of $\partial_n\mathcal{S}_{-k_j,k_j}$ (straight blue line) and $\partial_n\mathcal{I}_{(-k_i,k_i),(-k_j,k_j)}$ (dashed red line) in region II.}
 \label{meas-qpts}
 \end{centering}
\end{figure}

In Table \ref{table} we report the behavior at QPTs of the relevant partial derivatives of the studied quantities: Von Neumann entropy, quantum mutual information and negativity.

We start by taking as elementary subsystem the generic single mode $k_j$. At all the considered transitions we can recognize that the \emph{two-mode} correlations play a prominent role. Indeed, the correspondence of the divergencies of $\partial_x\mathcal{S}_{k_j}$ and $\partial_x\mathcal{I}_{k_j,-k_j}$ allows one to acknowledge the two-mode \emph{total} correlations between $k_j$ and $-k_j$ as relevant for the transitions II$\rightarrow$I and II$\rightarrow$III. The distinction between the latter resides in the fact that only at transition II$\rightarrow$III $\partial_u\mathcal{N}_{k_j,-k_j}$ is singular, suggesting that only in this case the two-mode \emph{quantum} correlations between the modes $-k_j$ and $k_j$ play a prominent role. We note, however, that the two-mode character of the transition II$\rightarrow$III is of a different nature with respect to the one seen in the direct lattice picture, where we observed a diverging entanglement range \cite{nota}. \\
This distinction in enforced by the relation between the negativity and the ODLRO and the fact that the derivatives of these two quantities display the same divergence at the transition II$\rightarrow$III. The comparison between these quantities and the mutual information between two arbitrary sites of the direct lattice discussed at the end of the previous section, allows once again to establish a link between the two-mode picture in $k$-space and the two-site picture in direct lattice. Indeed, due to the relation (\ref{Eq: Mutual_II_I_infty}) the divergence of $\partial_u \mathcal{I}_{|i-j|}^{II}$ at the transition II$\rightarrow$III, that allowed in \cite{AGM} to describe this transition as a two-point one, is in part determined by the divergence of the $\partial_u\mathcal{N}_{k_j,-k_j}$.\cite{nota2}

We now consider the elementary subsystem formed by the paired modes $(-k_j,k_j)$. The vanishing negativity $\mathcal{N}_{(-k_i,k_i),(-k_j,k_j)}=0$ shows that the \emph{two-pair} correlations measured by $\mathcal{I}_{(-k_i,k_i),(-k_j,k_j)}$ are always classical; this suggests the presence of just a \emph{multi-pair} structure of quantum correlations. The divergencies of $\partial_{x=u,n} \mathcal{S}_{-k_j,k_j}$ show that these correlations play an important, though different, role in both kind of transitions. The different  role can be described at the level of \emph{two-pair} correlations allowing for a classification of the QPTs into two classes: two-pair (transition II$\rightarrow$III) and multi-pair (transition II$\rightarrow$I). Indeed, as shown in table \ref{table} and in figure \ref{meas-qpts}, the singular behavior of $\partial_x \mathcal{I}_ {(-k_ik_i),(-k_jk_j)}$ and $\partial_x \mathcal{S}_{k_j,-k_j}$ coincides only at transition II$\rightarrow$III. As in the direct lattice picture, we can then conclude that also in $k$-space that the transition II$\rightarrow$III is characterized by the fact that correlations of two-point (two-pair) nature are the more sensible to the undergoing phase transition. On the contrary, the absence of a divergence of $\partial_x \mathcal{I}_{(-k_ik_i),(-k_jk_j)}$ suggests that correlations of multipartite (multi-pair) nature are more relevant at the transition II$\rightarrow$I. In the next section we will provide a further confirmation of such statement by directly evaluating a measure of multipartite correlations.

We now pass to the comparison of the singularities of the partial derivatives of the measures, as reported in Table \ref{table}, with those shown by the second partial derivatives of ground-state energy, (\ref{Energy Divergency II->I}), (\ref{Energy Divergency II->IV,II}). Interestingly, at transition II$\rightarrow$III all the computed quantities develop the same singularity of $\partial_u^2 E_{gs}(u,n)$. On the other hand, at transition II$\rightarrow$I $\partial_{x=u,n}^2 E_{gs}(u,n)$ exhibits a finite discontinuity and the same behavior is shown by both negativity and quantum mutual information, while the Von Neumann entropy is logarithmically divergent.

To summarize, it seems that when the transitions are signaled by divergencies the main role is played by ``two-point" correlations. Instead, in the case of  simple finite discontinuities the dominant role is played by the multipartite quantum correlations. In both cases, the appropriate derivatives of Von Neumann entropy diverges.

\begin{widetext}
\begin{table}[h]
\begin{center}
\begin{tabular}{|l|c|c|c|}
\hline
    & $\partial_x\mathcal{S}_{k_j}\;$,  $\partial_x\mathcal{S}_{k_j,-k_j}\;$,  $\partial_x\mathcal{I}_{k_j,-k_j}$ & $\partial_x\mathcal{N}_{k_j,-k_j} \;$, $\partial_x\mathcal{I}_{-k_jk_j,-k_ik_i}$ &\\
\hline
    $II\rightarrow I (x=u)$
    & $\log(u_c-u)$ &  FD & Multi\\
\hline
    $II\rightarrow I (x=n)$
    & $\log|n-n_c|$ & FD & Multi\\
\hline
    $II\rightarrow III (x=u)$
    &  $-1/\sqrt{u-u_c}$ & $-1/\sqrt{u-u_c}$ & Two\\
\hline
\end{tabular}
\end{center}
\caption{Behavior of the evaluated partial derivatives at the various QPTs (left column). 'FD' stands for Finite Discontinuity, 'Multi' for Multi-pair, while 'Two' stands for Two-pair.} \label{table}
\end{table}
\end{widetext}

We conclude this section by considering the half filling case. The latter has to be treated separately since no divergencies in the derivative of any measure of correlation can be observed at the transitions. This feature is a particular case of a more general phenomenon. As previously highlighted, in the TDL all measures of correlation depend on the average occupation number of each of the modes belonging to the set $K_{k>k_{N_s}}$. Therefore the curves in the $n,u$ plane defined by fixing $\average{n_{k_j}}$ to a given constant value can be considered as {\em iso-correlation curves} (an example of them is given in figure \ref{fig:phdiag}). Along such curves the correlations of any accessible subsystem do not change.
%({\bf in k-space!!! discuss how ODLRO changes along these curves, in particular what happens @ half-filling??}).
Consequently by crossing the boundary separating region II and III along these curves no divergence in the derivative of the correlations measures would be observed.\\
In particular, the condition $n=1$, i.e., the half filling case, defines a curve which is simply parallel to the $u$ axis. Thus, while $N_s$ -- that determines the number of available modes -- and $N_d$ can vary with $u$, $\average{n_{k_j}}$ is constant along this curve and all measures of correlations in region II are constant and fixed to the maximal achievable value (e.g., $\mathcal{S}_{k_j}=2$). In order words, in the $k$-space picture the transition between region II and III, while it is well described by all measures of correlations for $n\neq1$, at half filling (at variance with respect to the direct lattice case) and along the iso-correlation curves, it simply does not have any impact on the structure of the correlations of the state.

\subsection{Multipartite entanglement}\label{sec:result_multipartite}

In order to evaluate directly the multipartite part of the correlations and their behavior at QPTs, we consider the generalization of the Meyer-Wallach measure of multipartite entanglement $Q(\ket{\psi})$ \cite{MeyWalQ} to multi-qudit states introduced by Scott in \cite{ScottQ}:
\beq
    Q_{m,d} = \frac{d^m}{d^m-1}\left(1- \frac{1}{C^n_m}\sum_{\vec{i}} \mbox{tr} \rho^2_{\vec{i}}  \right)\; .
\label{Qmeas}
\eeq
Each multi index $\vec{i}= \{i_1,\dots,i_m\}$ identifies $m$ elementary subsystems e.g., momentum modes in our case, that compose a given principal subsystem $\mathcal{S}_m$ e.g., block of $m$ modes. The latter is characterized by its reduced density matrix $\rho_{\vec{i}}$. The factor $d^m/(d^m-1)$ normalizes the measure to one; $d$ is the dimension of the Hilbert space of each elementary subsystem and in the case of momentum modes it is equal to $4$.
$Q_{m,d}$ can be expressed as the average linear entropy of $\mathcal{S}_m$ and thus quantifies the average entanglement between blocks of momentum modes and the rest of the system. The more the size of the blocks is increased, the more $Q_{m,d}$ can capture and measure the multipartite correlations present in the system \cite{ScottQ,deOliveira}. In order to obtain a proper measure of the latter it is thus necessary to be able to compute the reduced density matrix of blocks of modes of arbitrary dimension.
While in the direct lattice this is possible only in region III \cite{salerno1}, in $k$-space one can generalize the results obtained in \cite{GA} and extend them to region II.
The extension is obtained again by recognizing that, in $k$-space, the role of the single particle with given spin $\sigma_{k_j}$ and momentum belonging to the set $K_{k_j \leq k_{N_s}}$ is to reduce the momentum spectrum available for the delocalization of the eta pairs to a subset of $L-N_s$ modes. One can thus derive the subsystems' density matrix  spectrum and, by implementing the definition of (\ref{Qmeas}) one has for $m=D$:
\bae
    Q_{D,4} &=& \left[1-{L-N_s\choose D}^{-1} \sum_{D_2=0}^D f(D_2)\mbox{Tr}(\rho_{D_1}^2)\mbox{Tr}(\rho_{D_2}^2) \right] \times \nonumber\\
    &\times&\frac{4^D}{4^D-1}    \label{KspaceMW}
\eae
where
\bae
    \mbox{Tr}(\rho_{D_1}^2) &=& \sum_{M=0}^{2D_1}{2 D_1\choose M} \left[\frac{{L-N_s-2D_1\choose N_d-M}}{{L-N_s \choose N_d}}\right]^2,\\
    \mbox{Tr}(\rho_{D_2}^2) &=& \sum_{\alpha=0}^{D_2}\left[{D_2\choose \alpha} \frac{{L-N_s-D_2\choose N_d-\alpha}}{{L-N_s \choose N_d}}\right]^2\;,\label{Tr-MW-kspace}
\eae
and
\bae
    f(D_2)&=&\frac{\prod_{i=0}^{D_1-1}(L-N_s-2i)}{D_1!} \times \nonumber\\ &\times& \frac{\prod_{j=0}^{D_2/2-1}(L-N_s-2D_1-2j)}{D_2!}\label{MW-kspace}
\eae
is the number of equivalent partitions of $D$ momentum modes into $D_1=D-D_2$ single modes $k_j$ and $D_2/2$ paired of modes $(-k_j,k_j)$.

\begin{figure}[!h]
    \begin{centering}
    \fbox{\includegraphics[height=7cm, width=8cm, viewport= 0 0 680
    550, clip]{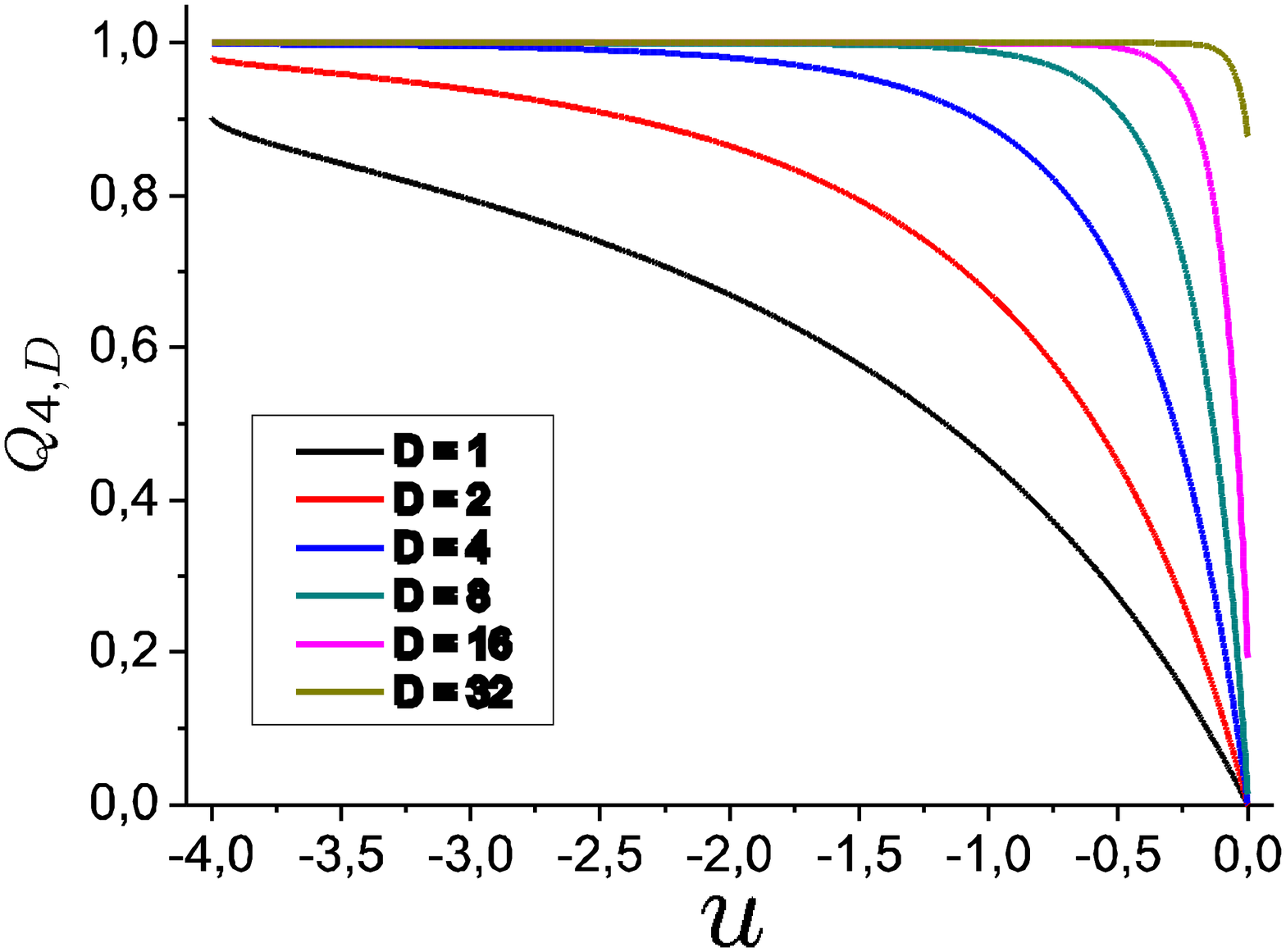}}
%    \includegraphics[height=7cm, width=8cm, viewport= 0 0 680
%    550, clip]{DerivMW.eps}}
%    \caption{$L=1000$, quarter filling, $D=1,2,4,8,16,32$ $Q$-measure (left) and its partial derivative (right).}
    \caption{Plot of the $Q$-measure for a chain of length $L=1000$ at quarter filling; $D=1,2,4,8,16,32$.}
    \label{Fig:KspaceMWRegII}
    \end{centering}
\end{figure}

In figure \ref{Fig:KspaceMWRegII} we plot the $Q$-measure (16) at quarter filling ( $L = 1000$, $N = 500$) as a function of the Coulomb interaction $u$ and for increasing values of the subsystem size $D$. The data confirm that – despite the reduced portion of $k$-space available for the construction of the eta pairs – the multipartite entanglement induced by their presence is of fundamental importance in the whole region II. Indeed, for all values of $u$, we observe an increasing saturation of $Q_{4,D}$ to its maximal value when increasing $D$ (for $D\ge 4$). Furthermore, for any dimension $D$ of the block of modes, $Q_{4,D}$ is  a monotonic decreasing function of $u$ and thus, since $n_d = 1/2(n-n_s) = 1/2(n-\arccos{(-u/4)}/\pi)$, it is a monotonic increasing function of the number $N_d$ of eta pairs.\\

The multipartite entanglement contribution at the various QPTs can be inferred by analyzing the behaviour of $Q_{4,D}$ at the critical points.
Again, it is only for $D\ge4$ that the correct qualitative behavior of $Q_{4,D}$ at large $D$ is captured: while at the the transition II$\rightarrow$III the curves smoothly reach the (constant) value exhibited by the measure in region III, the plot  clearly shows that $\partial_uQ_{4,D}$ is highly enhanced in proximity of the transition II$\rightarrow$I. The result, confirmed by numerical analysis, thus indicates the presence of a divergent behavior at large $D$.
This is a quantitative confirmation of the peculiar multipartite character of the transition II$\rightarrow$I: at that transition, though all correlations vanish, the multipartite ones change in a much faster way. Hence, the direct measure of multipartite correlations fully confirms the predictions of their role at QPTs, as obtained in sub-section \ref{sec:results_entqpt}. We emphasize that  a study at low values of $D$ -- like the one that would be feasible in the direct lattice picture -- would have suggested the opposite conclusion regarding the relevance of multipartite correlations at the transitions.

We finally note that the determination of the reduced density matrices $\rho_D$ for blocks of modes of arbitrary dimension $D$ allows for the evaluation also of the block entropy i.e., the Von Neumann entropy of $\rho_D$. This can  be done by extending the results obtained in \cite{GA}. This extension shows that even in the case analyzed in this paper the asymptotic behaviour of the block  entropy depends on the type of modes included in the block. In particular, the relevant result is that when the block is composed by $D/2$ paired modes only, the block entropy grows logarithmically with $D$. This typical growth shows on one hand the presence of multipartite entanglement and on the other hand the peculiar  multi-pair nature of this kind of entanglement.

\section{Conclusions}\label{sec:conclusions}

We investigated in momentum representation the structure of correlations of the ground state of an extended 1D Hubbard model, in relation to the undergoing QPTs. At variance with the standard real-space approach, the k-space representation allows in this case to explicitly derive the reduced density matrices of arbitrary subsystems, and thus to evaluate rigorously all bipartite and multipartite correlations present in the system.\\
A first interesting result is directly related with the presence of single electrons with given spin and momentum; the consequence of the presence of these particles is the limitation of the portion of momentum space available for the implementation of the  eta-paring mechanism. We show that this fact has various important implications in the structure and the behaviour of the correlations present in the ground state. In particular, we show that the phase space of the model can be characterized by the presence of \emph{iso-correlation} curves, i.e. curves in the phase space along which the value and the structure of the correlations does not change.\\
Due to the presence of the eta pairing mechanism, at a local level the existing network of correlations can be described in terms of two elementary building blocks: the generic single mode $k_j$ and the paired modes $(k_j,\,-k_j)$.
The identification of such elementary subsystems and the comparison of their (quantum) correlations with the results obtained in the direct lattice picture allows one to unveil the origin of the off-diagonal long-range order (ODLRO), whose value can be functionally expressed in terms of quantum correlations between two paired modes, i.e., negativity . As a consequence, we also show that the mutual information between two sites can be directly linked with the negativity between paired modes. \\
The identification of the above elementary subsystems has also allowed for the  classification of the QPTs in terms of two-pair or multi-pair correlations, depending on whether partial derivatives of quantum mutual information and Von Neumann entropy show the same singularity or not at a given QPT. These singularities can be related to those shown by second partial derivatives of the ground-state energy and, when possible, to the the singularities of measures of correlations derived in the direct lattice picture.\\
Finally, we have computed a proper direct measure of multipartite entanglement, the $Q$-generalization of the Meyer-Wallach measure \cite{MeyWalQ,ScottQ}
The latter requires the evaluation of the reduced density matrices relative to subsystems constituted by blocks of modes of arbitrary size $D$.  We have shown that, besides the different behaviour of the measure depending on the presence of paired/unpaired modes in the blocks \cite{GA}, as $D$ grows the partial derivative of $Q$ develops a singularity in correspondence with those QPTs at which the dominant correlations were conjectured to be of a multipartite nature. On the other hand in correspondence with the QPTs classified as "two-pair", by increasing $D$ the partial derivative of $Q$ becomes smooth.

The comparison between the structure of correlations in momentum space and their direct lattice counterpart turned out to be a useful method that allowed to unveil interesting relations between the correlations in the different pictures and to accurately characterize the ground state phase space and the quantum phase transitions occurring in a strongly correlated electron system. We expect that such method can be fruitfully extended to other many body complex systems.

%%%%%%%%%%%%%%%%%%%%%%%%%

\begin{thebibliography}{0}
\bibitem{nota}
The range of entanglement is defined as the maximum distance between two subsystems at which a given measure of entanglement (e.g. concurrence or negativity) is different from zero. In \cite{AGM} it was shown that in correspondence with QPTs classified as two-point the range of entanglement is diverging, while when approaching a multipartite QPT it remains finite.
\end{thebibliography}
\end{document}